\documentclass{article}

\RequirePackage{etex}

\usepackage[utf8]{inputenc}
\usepackage{todonotes}
\usepackage[T1]{fontenc}
\usepackage{algorithm}
\usepackage{hyperref}
\usepackage{authblk}
\usepackage{xspace}

\usepackage[
    n,
	lambda,
	operators,
	advantage,
	sets,
	adversary,
	landau,
	probability,
	notions,	
	logic,
	ff,
	mm,
	primitives,
	events,
	complexity,
	oracles,
	asymptotics,
	keys]{cryptocode}
	
\usepackage{geometry}
\newcommand{\hygiea}{\textsc{Hygiea}\xspace}
\newcommand{\gb}{Governing Body\xspace}
\newcommand{\ci}{Certificate Issuer\xspace}
\newcommand{\cis}{Certificate Issuers\xspace}
\newcommand{\ch}{Certificate Holder\xspace}
\newcommand{\chs}{Certificate Holders\xspace}
\newcommand{\vf}{Verifier\xspace}
\newcommand{\vfs}{Verifiers\xspace}

\title{\hygiea: A secure, smart, privacy-preserving and interoperable
Blockchain solution for the Covid-19 pandemic}
\author[1]{Sofia Maria Dima}
\author[2]{Alexandros Hasikos}
\author[3]{Stylianos Kampakis}
\author[4]{Theodosis Mourouzis}
\author[5]{Andreas Papageorgiou}
\affil[2]{UPF Barcelona}
\affil[1, 2, 3, 4, 5]{Electi Consulting Ltd}
\affil[3, 4]{UCL CBT}
\affil[4]{LBS}

\date{}

\begin{document}

\maketitle

\begin{abstract}
In this article we present \hygiea, an end-to-end blockchain-based solution for the Covid-19 pandemic. \hygiea has two main objectives. The first is to allow governments to issue Covid-19 related certificates to citizens that can be verified by designated verifiers to ensure safer workplaces. The second is to provide the necessary tools to experts and decision makers to better understand the impact of the pandemic through statistical models built on top of the data collected by the platform. This work covers all steps of the certificate issuance, verification and revocation cycles with well-defined roles for all stakeholders. We also propose a governance model that is implemented via smart contracts ensuring security, transparency and auditability. Finally, we propose techniques for deriving statistical models that can be used by decision makers. 
\end{abstract}

\section{Introduction}
The COVID-19 pandemic has caused major disruptions in society, economy, business and healthcare. At the time of this writing, the pandemic caused death to millions of people \cite{JohnHopkinsCovid}. Governments, along with the World Health Organisation and other experts, had to propose and enforce tough measures such as social distancing, masks and lock-downs in an attempt to contain the virus and offload the health system. 

The scientific community, pharmaceutical companies and regulators are working hard to fabricate and license vaccines that will be provided to people in order to increase the resistance against the virus' symptoms and as result contain the spread. So far, the process has been proven hard for all stakeholders given the time constraints. In addition, the rapid spread of the virus has generated an enormous demand for vaccines, a fact that imposes additional challenges to the supply chain.

While the healthcare industry is working hard towards the manufacture and supply of vaccines there is a need for an efficient mechanism to collect data that will assist in better understanding and analysis of the virus. Examples of data can be type of symptoms, intensity, antibodies and geographic location. In addition, analysis of the data will significantly improve the quality of actions taken by decision makers and authorities as they attempt to contain the spread and return back to normality. Another aspect of data collection is the standardisation of the format as well as the protocols followed to collect them and make them available to the relevant parties while considering privacy.

The transition to normality is not expected to happen instantly, thus the term ``new normality'' has been introduced which will require people to continue keeping precautions such as social distancing in order to contain the spread. In addition, travelling, commuting and working will require people to present various types of proof in a form of certificates to verify their health condition. Examples of these certificates are COVID-19 negative tests, vaccination and recovery. Although this ``new normality'' is considered inconvenient and has been heavily criticised\cite{Halpin2020}\cite{Voo2020} in terms of its ethical, social and practical consequences, several countries\cite{VaccinePassportMonitor} across the globe are considering a digital solution that will speed-up the recovery from the pandemic.

In an attempt to address the consequences of the pandemic, the European Commission has published a call for proposals for a Digital Green Certificate\cite{EUDigitalGreenCertificate} which will facilitate safe and free movement during the COVID-19 pandemic within the EU. According to the proposal, the Digital Green Certificate (DGC) system must cover three types of certificates. Vaccination, COVID-19 test and recovery. All citizens of the EU are eligible to be issued certificates and the certificate will include only a minimum set of information necessary to confirm the holder's status.

\section{Related Work}
In \cite{Hasan2020} authors propose a blockchain-based solution that implements digital medical passports and immunity certificates for COVID-19 test takers. The solution is based on the public Ethereum blockchain and utilizes smart contracts, IPFS and proxy re-encryption techniques. The architecture of the system includes both on-chain and off-chain entities that are governed through a central trusted authority. Certificates of the users are stored encrypted in a distributed file system (IPFS) and users can reveal the contents of the certificates to interested parties using a technique called ``proxy re-encryption'' through a trusted entity. In \cite{Eisenstadt2020} authors propose an immunity passport system that is based on W3C's Verifiable Credentials\cite{VerifiableCredentials} standard, the Solid\cite{Solid} decentralised data platform and a consortium blockchain. The system covers three roles, namely. Issuers, holders and verifiers who are onboarded to and authenticated by system. Issuers can issue certificates to holders in a form of verifiable credentials that are signed by both the issuer and the holder and stored on the holder's Solid Pod (wallet). A hash of the certificate is also stored on the consortium based blockchain for verification. Verifiers verify both the digital signatures and the hash on the claimed certificates. In \cite{Untung2020} authors propose a simple COVID-19 test certification framework where, similarly to \cite{Eisenstadt2020}, certificate issuers, holders and verifiers are onboarded to a platform. Holders are assigned a decentralised identity that is used by issuers to issue COVID-19 related certificates. The certificate takes the form of a hash and is stored on a consortium blockchain. Verifiers can verify certificates by comparing the hash found on the blockchain and the hash of the actual medical information presented by users. In \cite{secureabc2020} authors propose a protocol for the issuance and verification of antibody certificates for COVID-19. Although this proposal does not involve the use of blockchain, it still achieves all the necessary functionality of secure and privacy preserving issuance and verification of certificates. In addition, it covers both paper and digital certificates. In~\cite{aydar2020blockchain} authors propose a blockchain-based framework for digital identity verification, record attestation and record sharing using verifiable credentials in a decentralised setup. In their work, they present as a use-case the COVID-19 certificates problem. In \cite{Ahmad2020} authors discuss several applications of blockchain against the COVID-19 pandemic. One of the applications discussed in their work is the vaccination certificates and immunity passports for which they give a high level architecture of the proposed system. The system, similarly to \cite{Hasan2020}, uses the Ethereum blockchain, smart contracts, IPFS and proxy re-encryption. In \cite{hernandez2021sharing} authors consider a setup in which all EU countries participate. The system is based on decentralised identifiers and verifiable credentials for issuance and verification of vaccine certificates. The Hyperledger\cite{Hyperledger} blockchain is used for storing hashes of the certificates. In this work, authors provide implementation details with performance metrics. In \cite{chaudhari2021framework} authors address the problem of user authentication by proposing the use of biometrics and specifically the extraction of iris templates and a hashing mechanism to store them on the blockchain. The system uses a permissioned blockhain and digital signatures for certificate authentication.

\section{Our Contribution}
In this work, we propose a multi-purpose platform named \hygiea. \hygiea is a platform designed for the issuance and verification of Covid-19 certificates that combines blockchain, smart contracts and cryptography and is realistic to implement in practice, flexible to adapt in various scenarios, secure for all participants and respects user's privacy.

The main purpose of \hygiea is to provide an ecosystem on which regulators, certificate issuers, certificate holders and verifiers can operate on, in order to streamline the process of issuance and verification of Covid-19 certificates. In addition, through the use of \hygiea we aim to prevent falsification of certificates\cite{Europol} and increase trust through decentralization by using the blockchain. Another goal of our system is to assist the health community to better understand the virus by providing a way of collecting anonymized data from patients and deriving analytics.

The system is built on top of Ethereum deployed in a permissioned setup which enables to define all the business logic encoded in the form of smart contracts. In addition, all the stakeholders can co-exist in the system while having well-defined processes and responsibilities. The permissioned setup also enables both trusted and non-trusted entities to operate in the same environment. The efficiency and performance mostly found on public blockchains is solved by using Proof-Of-Authority as a consensus mechanism.

The governance rules are encoded in the form of smart contracts. A trusted entity, namely the governing body, is the owner of the governance smart contract and can, for example, onboard other entities which have a predefined list of operations that are allowed to perform. The integrity of the operations on the blockchain as well as the data are secured through the use of solid cryptographic operations already provided by Ethereum.

\hygiea is an end-to-end proposal meaning that it covers all steps of the process. Starting from the governance of the system, specification of the stakeholders and their roles, protocols for the issuance and verification of certificates as well as certificate revocation. To our knowledge, \hygiea is the first blockchain-based system that conforms with EU's proposal for the implementation of the Digital Green Certificate. In fact, \hygiea can operate both together with and independently from with the European Union's (EU) DGC. Another advantage of \hygiea is that it provides a tool for scientists and decision makers to create analytics in order to combat the virus more effectively. The modular design of the system can include further stakeholders such as drug, vaccine and medical equipment manufactures in order to address supply chain challenges that occurred during the pandemic.

\section{\hygiea components and technologies}
\label{sec:components and technologies}
In this section, we will introduce all the terms necessary and components to understand all the operations and interactions in the \hygiea protocol. 

\subsection{Blockchain}
\label{subsec:blockchain}
Blockchain has been famously introduced through the Bitcoin cryptocurrency \cite{bitcoin2009} in 2009. Blockchain is a distributed ledger technology (DLT) that runs on multiple nodes in a peer-to-peer (P2P) network and is used to record and secure transactions. Changes on the state of the distributed ledger are achieved through a consensus protocol which ensures that all participating nodes in the network agree on what the new state is. There exist several consensus protocols depending on the application and each of them has pros and cons but the idea behind is the same (a mechanism to help non-trusted nodes come into consensus). In the case of Bitcoin the consensus protocol used is called Proof-of-Work (PoW). In this protocol, the participating nodes in the network are required to solve a hard cryptographic problem that requires computational resources(a.k.a mining). The miner who manages to solve the problem is the one that will update the state (or create a new block) of the blockchain and get rewarded. Apart from PoW several other consensus protocols, namely Proof-of-Stake (PoS) and Proof-of-Authority (PoA) are being used to address the increased costs and scalability limitations of PoW. In a PoS setup, stakeholders vote with their stakes meaning that every account has a certain chance per second of generating a valid block and this chance is proportional to the account’s balance. In PoA, which is a less decentralised architecture, only certain validators can create new blocks. Validators can be selected according to the system's governance rules. A resulting property of blockchain is that all transactions on the ledger are tamper-proof and easily verifiable making it ideal for storing information requiring resistance to tampering. 

Certain blockchains, such as Ethereum and the Hyperledger family, can be deployed in a private or permissioned setup. Permissioned blockchains, as opposed to public, allow only a number of trusted network nodes to exist in the network and are usually deployed in business applications where consensus can be achieved based on authority(using validator nodes). Since trust is an implied property in such a setup, consensus can be achieved using a more scalable and efficient protocol such as PoA. Another reason for deploying private blockchain is the requirement of data confidentiality in the network as well as reduced costs and increased performance. In a private setup, miners can be authorised nodes that are responsible for validating transactions and changing the state of the distributed ledger.

In this work we will only consider the Ethereum blockchain deployed in a permissioned setup and using the Proof-Of-Authority consensus protocol. The reasons behind the choice are the combination of trusted and non-trusted stakeholders in the system, the reduced cost and the performance requirements of the application. 

The use of a public blockchain would involve significant costs for the deployment and operation of the smart contracts as well as performance bottlenecks resulting from the slow mining process.

\subsection{Smart Contracts}
\label{subsec:smart contracts}
Smart contracts in the context of the blockchain technology simply refers to computer code that runs on the blockchain and can be used to build custom decentralised applications (DApps). A rough analogy of smart contracts in the context of relational databases are stored procedures where operations can be applied on the data stored. 

Smart contracts allow us to run programs that can perform actions by following rules when certain events are triggered. Ethereum \cite{wood2014ethereum} was the first blockchain to utilise smart contracts that allowed people to build applications by providing a dedicated programming language called Solidity\cite{Solidity}.

There are two parts in a smart contract. The first one is its state which is all the data that is stored and the second are the functions that can be used to change the state. A powerful feature of smart contracts is that all state changes are recorded on the blockchain, making it transparent and auditable. In addition, all transactions are cryptographically signed and secured using digital signatures.

A smart contract, once deployed, gets assigned with an address which can be used as a reference for interaction. Interaction with a smart contract can be either updating or reading the state. Updating refers to calling a function which changes the values of member variables of the code and reading refers to viewing the values of the member variables. Both actions can be achieved through the Application Binary Interface (ABI) which is one of the artifacts generated when compiling the source code of a smart contract. 

For transactions that update the state of the smart contract, there is an associated cost which reflects the effort required to run a certain piece of code as well as the amount of memory required for storage. In the case of permissioned blockchains, this can be omitted since the cost of operation can be distributed over the members of the consortium. 

A smart contract can be used in numerous applications such as the exchange digital assets (cryptocurrencies, non-fungible tokens), Decentralized Finance (DeFi), online service subscriptions, electronic voting, healthcare\cite{BlockchainInHealthcare} and many others. In a decentralized system in which there is no trust, smart contracts are deployed to provide transparency in terms of its governance and functionality.
\subsection{Digital Signatures}
\label{subsec:digital signatures}
A digital signature is an asymmetric cryptographic scheme that allows the verification of authenticity of a digital piece of data or document. Suppose we have Alice who has a document and wishes to generate a piece of information that can be used to prove that Alice herself approves of the document, similarly to the signatures placed in ordinary paper. Anyone in possession of the document and the associated digital signature can verify that indeed Alice was the one who generated the signature on that particular document. 

\begin{figure}
    \centering
    \includegraphics[scale=0.8]{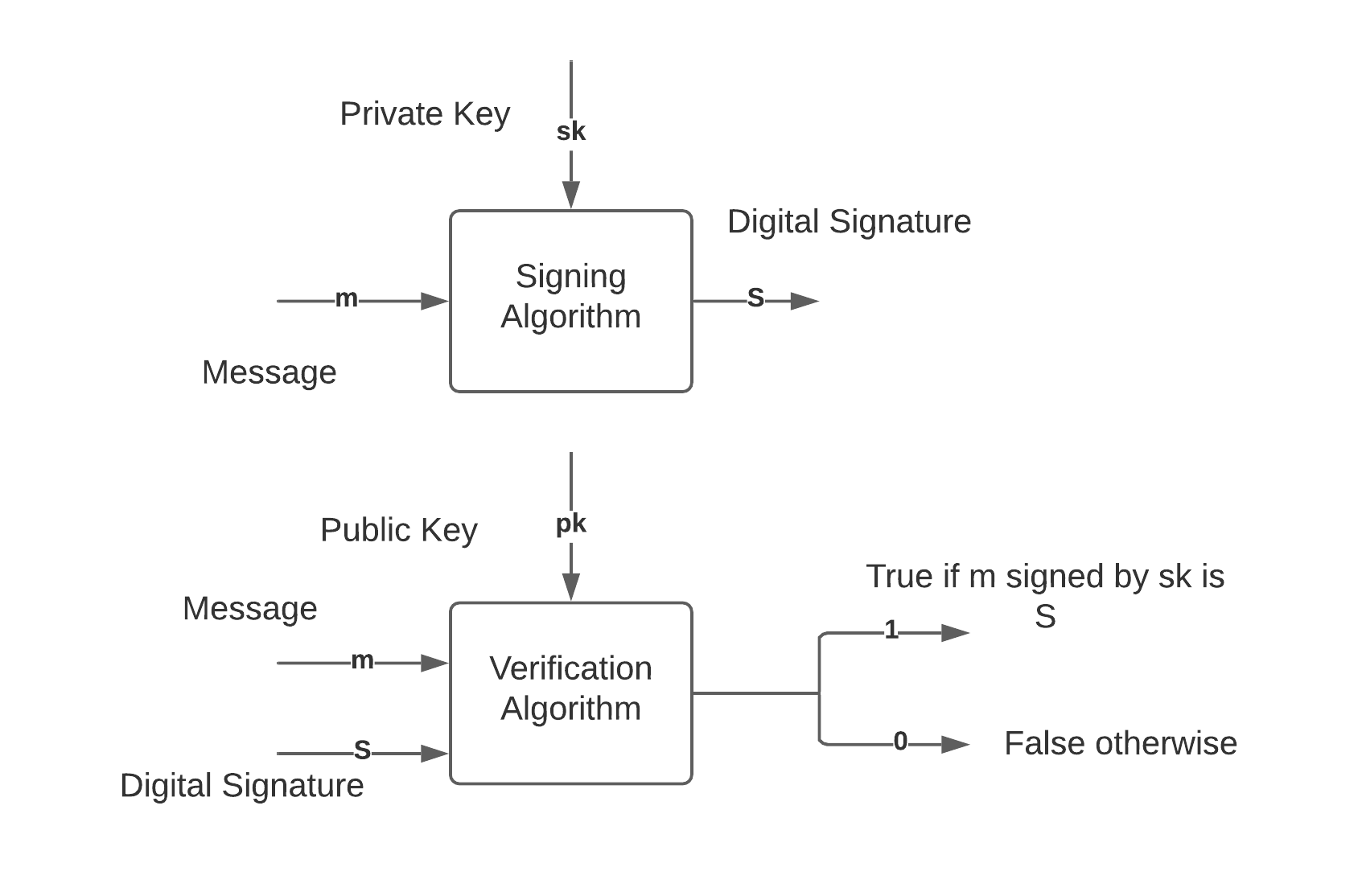}
    \caption{Digital Signatures}
    \label{fig:digital signature}
\end{figure}

Figure \ref{fig:digital signature} depicts how a digital signature scheme works. A typical digital signature scheme involves the use of a cryptographic keypair consisting of the secret key $\sk$, that must be kept private to the owner, and a public key $\pk$, which can be publicly available, a message $m$, a signature algorithm $\sign$ and a verifying algorithm $\verify$.
 
 The owner of the secret key $\sk$ can perform a signing operation $\sign_{\sk}(m)$ with inputs the secret key $\sk$ and the message $m$ and output a signature $s$. Verifiers can then use the verifying algorithm $\verify$ with inputs the signature $s$, the message $m$ and the public key $pk$ such that $\verify_{\pk}(s, m)$ and obtain the output which can be 0 or 1.
 
The Ethereum blockchain uses the Elliptic Curve Digital Signature Algorithm (ECDSA) with secp256k1\cite{secp256k1} parameters. ECDSA is a direct analogue of the Digital Signature Algorithm (DSA)\cite{DSA} and its security relies on the hardness of the elliptic curve discrete logarithm problem. ECDSA is a very efficient digital signature scheme that is in widespread use especially in situations where the signature size is important. 

\subsection{Identification schemes}
\label{subsec:identification schemes}

\begin{figure}
    \centering
    \includegraphics[scale=0.4]{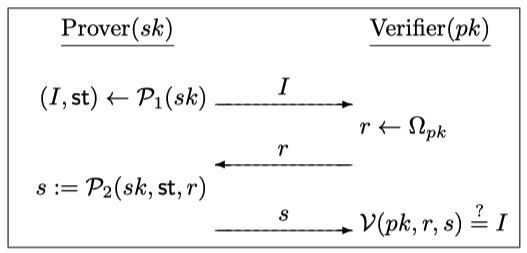}
    \caption{Three round identification protocol}
    \label{fig:three round identification protocol}
\end{figure}

\begin{figure}
    \centering
    \includegraphics[scale=0.4]{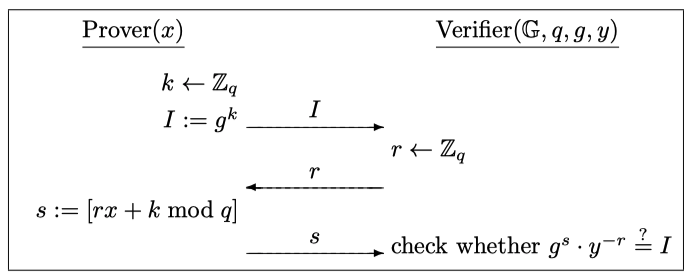}
    \caption{Schnorr's three-round identification scheme}
    \label{fig:schnorr's three-round identification scheme}
\end{figure}

An identification scheme is an interactive protocol that allows a party (the prover) to prove its identity to another (the verifier). The class of identification schemes that are of interest are the public-key based ones where the prover is not required to share any private information in order to prove their identity to the verifier. The execution of such a scheme requires that the two parties interact in order to exchange messages. An example three-round identification scheme is depicted in figure \ref{fig:three round identification protocol} and the five round in Figure \ref{fig:schnorr's three-round identification scheme}. In the first step of the protocol, the prover uses an algorithm $\mathcal{P}_1$ and the secret key to obtain an initial message $I$, which is sent to the verifier, along with some state $\text{st}$. The verifier generates a challenge $r$ from some set and sends it to the prover. The prover runs function $\mathcal{P}_2$ with inputs being the secret key, the state and the challenge. The output $s$ is sent to the verifier who runs function $\mathcal{V}$ and checks if it matches the initial message $I$.

Schnorr's\cite{schnorr89} identification scheme is a three-round scheme based on the hardness of the discrete logarithm problem similarly to ECDSA. In \hygiea, Schnorr's identification scheme is used for the authentication during issuance and verification of certificates. Figure \ref{fig:schnorr's three-round identification scheme} depicts the execution of the protocol.

%

\section{Stakeholders}
\label{subsec:stakeholders}
In this section we will explain in detail each stakeholder in the system as well as the roles associated.

\subsection{\gb}
\label{subsubsec:governing body}
The \gb is the stakeholder responsible for the governance of the system. In practice, the \gb can be represented by the Ministry of Heath (MoH) or the National Health System (NHS). The existence of a trusted entity simplifies governance and gives additional assurance to the security of the system.

 We identified the following responsibilities of the \gb within the system:
\begin{itemize}
	\item \textbf{Set up the blockchain consortium}: The \gb acts as a trusted entity within the system. The governance of the consortium is specified in the form of smart contracts and deployed on the blockchain. Participating entities such as \cis and \vfs must comply with the governance rules in order to have valid status in the system.
    \item \textbf{Run a validator node:} The \gb is in control of a validator node within the system. As the name implies, the validator node has the authority to validate transactions on the blockchain. The transactions will be submitted by other transaction nodes participating in the network according to the governance rules of the network.
    \item \textbf{Register \cis:} \cis (see section \ref{subsubsec:certificate issuers}) are entities that are identified by a unique public address. The \gb can add \cis in a registry that is encoded in a smart contract and must be publicly accessible by all participants in the system including the \chs. The registration of a \ci consists of both an on-chain and off-chain process whereby it must be ensured by the \gb that the particular \ci is eligible for participation in the system. Apart from the registration process the \gb also designates which types of certificates a given \ci is allowed to issue (specified by the governance rules).
    \item \textbf{Register \vfs:} \vfs (see section \ref{subsubsec:verifiers}) are also required to be onboarded to the system. The registration process is identical to the \ci and there exists a public registry on which all approved \vfs are listed. Similarly to \cis, \vfs are identified by a unique public address.
    \item \textbf{Change access rights to \cis and \vfs:} The \gb has the right to revoke or change access rights to any misbehaving \ci or \vf. A revocation of rights simply means that a \ci will not be able to issue certificates or a \vf will not be able to verify any certificates. In the case of revocation of a given entity, the \gb updates the corresponding registry by removing the public address from the list. Any change to the state is transparent as it is an inherent property of the blockchain.
    \item \textbf{Revoke certificates:} The \gb will have the functionality to revoke certificates belonging to certificate holders. This means that the holders who claim revoked certificates will not be able to pass the verification process. Certificate revocation can be triggered in the case where there was an accidental issuance or due to a faulty batch of test equipment.
\end{itemize}

\subsection{\cis}
\label{subsubsec:certificate issuers}
\cis in \hygiea are entities that are authenticated and authorised by the \gb to issue certificates to holders. \cis can be Hospitals, Clinics, Test Labs, GPs or any other eligible entity. Each \ci will be given access to a DApp from which they will be able to issue certificates. A \ci must generate their own cryptographic key pair which must be stored in a dedicated wallet and will be used for signing and verification of issued certificates. A \ci can issue certificates to holders if and only if its public address has been registered by the \gb (its identity exists on the public registry on the blockchain). In summary, \cis:
\begin{itemize}
    \item \textbf{Authenticate \chs:} A \ch (see section \ref{subsubsec:certificate holders}) visiting a \ci in person needs to be authenticated both physically and digitally. Authentication is required to prevent malicious \chs from being issued certificates that are not entitled (impersonation attacks). 
    \item \textbf{Issue certificates:} \cis can issue certificates to  \chs. Certificates are issued in the form of smart contracts and submitted as transactions to the blockchain through designated transactions nodes. The validator node, belonging to the \gb, is responsible for validating the transaction and update the state of the blockchain. Each certificate contains, apart from the necessary data fields, a digital signature of the issuer. Note that \cis whose access has been revoked will not be able to issue certificates. 
    \item \textbf{Data collection:} Certain \cis can be healthcare providers who can perform medical examinations. In this case, given a person's consent, anonymous data can be collected for further analysis. The anonymised data will be stored in a special repository in which authorised entities will be granted access. The data collection process will be done through the same application used for issuing certificates.
\end{itemize}

\subsection{\vfs}
\label{subsubsec:verifiers}
\vf can be any official or unofficial entity participating in the system and has been approved by the \gb. \vfs need to be registered by the \gb in a similar fashion to \cis and are identified by a public address. \vfs will have to generate their own cryptographic keys and store them into a dedicated wallet. In \hygiea, two classes of \vfs are supported. The first class is the one that is allowed to change the state of the blockchain and the second is the one that can only read the blockchain. The main functionalities of \vfs are:

\begin{itemize}
	\item \textbf{Verify certificates:} Certificates claimed by \chs to \vfs in a form of a smart contract that exists on the blockchain. \vfs are required to ensure the authenticity of the certificate by using standard cryptographic techniques such as hash functions and digital signatures. Each certificate contains certain data, depending on the type, that are digitally signed by the issuer who is a trusted entity in the system. \vfs need to ensure that the particular certificate has not been revoked, the issuer's public identity exists in the registry of the \gb and finally that digital signature is verified.
	\item \textbf{Update the blockchain:} Certain \vfs are required to log when a particular certificate has been verified. Whenever a user claims a certificate, the \vf must update the blockchain with the a timestamp of the verification. An analogy is the process where a citizen enters another country and a stamp is issued on the passport.
	\item \textbf{Authenticate \chs:} \vfs must authenticate both the physical and digital identities to minimise the possibility of a forgery. Authentication processes will be presented in later sections.
\end{itemize}

\subsection{\chs}
\label{subsubsec:certificate holders}
\chs in \hygiea are defined as all physical entities with an associated digital identity that are entitled to be issued a certificate. Holders can participate in the system by using a mobile application that can be used to store their cryptographic keys and the addresses of their certificates. Each \chs will be responsible for the generation and storage of their keys and no other stakeholder will have any access to them.

\chs will:
\begin{itemize}
    \item \textbf{Generate and store keys:} Holders are responsible for creating and storing their own cryptographic keys. The keys associate themselves with their digital identity and this is required for authentication by both \cis and \vfs. The keys and the references to the certificates on the blockchain are held in a dedicated mobile application.
    \item \textbf{Authenticate \cis and \vfs:} Approved \cis and \vfs exist in a smart contract published on the blockchain by the \gb. \chs are required to authenticate \cis and \vfs in order to ensure that they follow the \gb guidelines and that their access has not been revoked before to issuance or verification of a certificate.
    \item \textbf{Proving their identity:} \chs are required to present their certificates to the \vfs. In addition, they are required to prove their identity in order to convince the \vfs that the certificate presented belongs to them. This is achieved through standard cryptographic operations such as hash functions and digital signatures.
\end{itemize}


\section{\hygiea}
\label{sec:hygiea}

\subsection{Blockchain Network}
\label{subsec:network setup}
\begin{figure}
    \centering
    \includegraphics[scale=0.33]{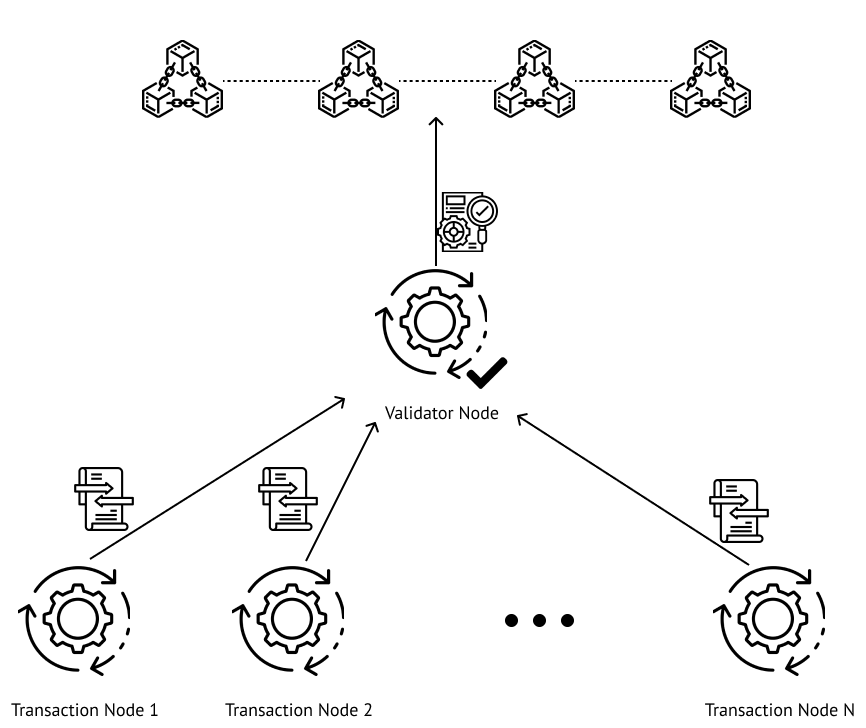}
    \caption{PoA Blockchain Network}
    \label{fig:blockchain network}
\end{figure}

\hygiea uses a private Ethereum blockchain running the PoA consensus protocol. PoA is a consensus algorithm that is practical, efficient, scalable and it is best suitable for private blockchain. In a PoA setup, a validator which is considered as trustworthy is chosen based on its identity and reputation as opposed to a Proof-of-stake(PoS) where validators are staking their coins. The number of validators is limited and must be verified by all participants of the system. The rest of participants run a transaction node which simply can submit transactions that are validated by the validator node. The formation of the consortium can be rearranged to accommodate different requirements for validator and transaction nodes. For example, consider a scenario in which there is more than one validator or the validator is determined by vote majority from the participants. Figure \ref{fig:blockchain network} depicts the network architecture.

In \hygiea, the validator node is run by the \gb and is responsible for validating transactions and appending them to the blockchain. The \gb is a trustworthy entity within the system and is publicly verified.  

Transaction nodes, run by \cis and \vfs, are nodes that are used to submit transactions which are validated by the validator of the consortium. 

In private PoA network such as the one we are considering for this application, the concept of \textit{gas} does not apply. Transaction nodes can submit their transactions without having to worry about processing and memory as the network is run privately on powerful cloud infrastructure. 

Another inherent property of the private network is privacy. All the data existing on the private blockchain are not exposed to the public since access control is enforced through the design. This makes the private blockchain more attractive to the applications where private or medical data might be stored.

\subsection{Smart Contracts}
\label{subsec:smart contracts}

\subsubsection{Governance}
\label{subsubsec:governance smart contract}
Governance is one of the most important aspects of \hygiea as it formalises and secures all the processes for the registration of the stakeholders, issuance and verification of the certificates. All the governance rules are encoded in a smart contract that is owned and deployed by the \gb. 

The primary purpose of the governance smart contract provides a mechanism for the \gb to register \cis and specify their rights within the system. Each \ci generates its own Ethereum keys that make them identifiable and unique within the system. The registration process simply means that the \gb, adds the public key of a given \ci to a list within the smart contract and associates the respective attributes:

\begin{itemize}
	\item \textbf{Country:} The country in which the issuer is based in. 
	\item \textbf{Name:} Name of the issuer in a human readable form.
	\item \textbf{Id:} The identifier of the issuer.
	\item \textbf{Allowed certificate types:} Types of certificates (Vaccine, Recovery, Test) the issuer is allowed to issue. The types are set by the \gb.
	\item \textbf{Valid from:} The timestamp from which the issuer is allowed to start issuing certificates.
	\item \textbf{Status:} The status(Active, Inactive) of the issuer dictates whether it is able to issue certificates. The status can be set and changed by the \gb.
	\item \textbf{Public key:} The public key of the issuer is its identity on the blockchain. All certificates issued by a given issuer will include this public key.
	\item \textbf{Signature:} All the above fields are concatenated and signed with the private key of the \gb. The signature can be publicly verified to ensure that only the \gb performed the registration.
\end{itemize}

The list as well as the attributes can only be modified by the \gb. This is ensured through the blockchain as the respective functions allowing the modification of the data structures can only be executed through transactions signed by the private key of the \gb. Another important feature of the data structure holding the attributes of \cis is  the signature. All the data fields related to a given \ci are cryptographically signed by the \gb thus providing an extra attestation.

A similar approach is followed for the \vfs. A certificate verifier as discussed in section \ref{subsubsec:verifiers} needs to be approved by the \gb. Although a \vf is not entitled to issue certificates, it can also submit transactions that log when and where a particular certificate has been verified.

The second purpose of the governance smart contract is to become a trustful registry for certificate holders and verifiers. The address of the governance smart contract must be publicly visible for:
\begin{itemize}
	\item \textbf{Certificate holders:} A certificate holder visiting a test lab, hospital or clinic has to be ensured that the issuer is indeed verified by the \gb and has a valid status. The governance smart contract originating from a trustworthy organisation can provide the required trust to the holder. Same applies for \vfs. A \ch needs to also trust whoever is claiming to be authorised to verify the certificate presented.
	\item \textbf{\vfs:} A verifier must know if the issuer of a certificate belongs to the list of approved and active issuers in order to verify a certificate. 
\end{itemize}

\subsubsection{Certificate Factory}
\label{subsubsec:certificate factory smart contract}

The certificate factory smart contract is owned by the \gb and is designed to provide templates to the various certificate types supported within \hygiea. The rationale behind this smart contract follows the factory design pattern\cite{designpatterns1995} from software engineering principles.

The data fields as well as the functions that must be present in a certificate are templated and \cis eligible to issue certificates to holders, use the certificate factory smart contract to issue certificates. In addition, the smart contract implements controls to prevent \cis with invalid status from issuing certificates as well as other checks for input sanitisation.

The addresses of all certificates issued are recorded into data structures for the \gb to derive analytics as to how many certificates have been issued and what is their address on the blockchain. Certificate holders use the certificate factory smart contract to look up for their certificates and add them to their wallet.

The benefit of the certificate factory smart contract is that it serves as a tool for both certificate issuers, holders and verifiers as it contains all the necessary information. The integrity of the information is ensured through the blockchain and it is ensured that at any given moment all stakeholders have a common and identical state of the data.

\subsubsection{Certificate}
\label{subsubsec:certificate}

Each holder within \hygiea will be issued multiple certificates which he can store on his wallet. A certificate in the context of \hygiea is a smart contract that has been designed to include the necessary information and functionality in order to prove possession and verify the authenticity of the data.

The purpose of the certificate smart contract is to encode the business logic of the certificates supported. Typically, what matters in a certificate is its status (whether valid or not) and depending on the \gb's specification it can be set accordingly. For this use case the validity of a certificate smart contract depends on:
\begin{itemize}
	\item \textbf{Expiration date:} If the expiration date of certificate has passed then the state of the smart contract becomes invalid. The exact data is set during the creation of the certificate and cannot be altered in the future.
	\item \textbf{Revocation:} The \gb has the authority to change the status of a certificate to be revoked. This means that only a transaction signed with the \gb's private key can alter the state. In such a case then the status of the certificate is invalid.
	\item \textbf{Invalid signature:} If the signature included in the certificate smart contract does not belong to any of the approved \cis then it cannot be accepted by a \vf.
\end{itemize}

Certificate smart contracts hold the following data fields:
\begin{itemize}
	\item \textbf{Personal identifier:} According to the identity binding mechanism used, this field will hold all the information relevant to authenticate the physical identity of a user.
	\item \textbf{Certificate type:} This can be any of Vaccine, Test or Recovery.
	\item \textbf{Issuance date:} The timestamp which the certificate was issued.
	\item \textbf{Valid from date:} The vaccination certificate becomes valid three weeks from the day it is issued. This field specifies the timestamp from which the certificate becomes valid. If the holder attempts to verify the certificate before that date it will get rejected.
	\item \textbf{\ci address:} The public key of the issuer who issued the certificate.
	\item \textbf{Governance smart contract address:} The address of the governance smart contract. This will be used in case of revocation where only the \gb will be allowed to perform this action.
	\item \textbf{Status:} This is a flag to specify whether a certificate has been revoked.
	\item \textbf{Signature:} All the above fields except from the status are concatenated and signed using the private key of the issuer.
\end{itemize}

\noindent \hygiea supports 3 types of certificates:
\begin{enumerate}
	\item \textbf{Vaccination:} Issued to citizens who received at least one dose of a Covid-19 vaccine.
	\item \textbf{Recovery:} Issued to citizens who recovered from Covid-19.
	\item \textbf{Test:} Issued to citizens who tested positive or negative to Covid-19.
\end{enumerate}

\subsection{\ci Registration}
\label{subsec:certificate issuer registration}
\begin{figure}
    \centering
    \includegraphics[scale=0.33]{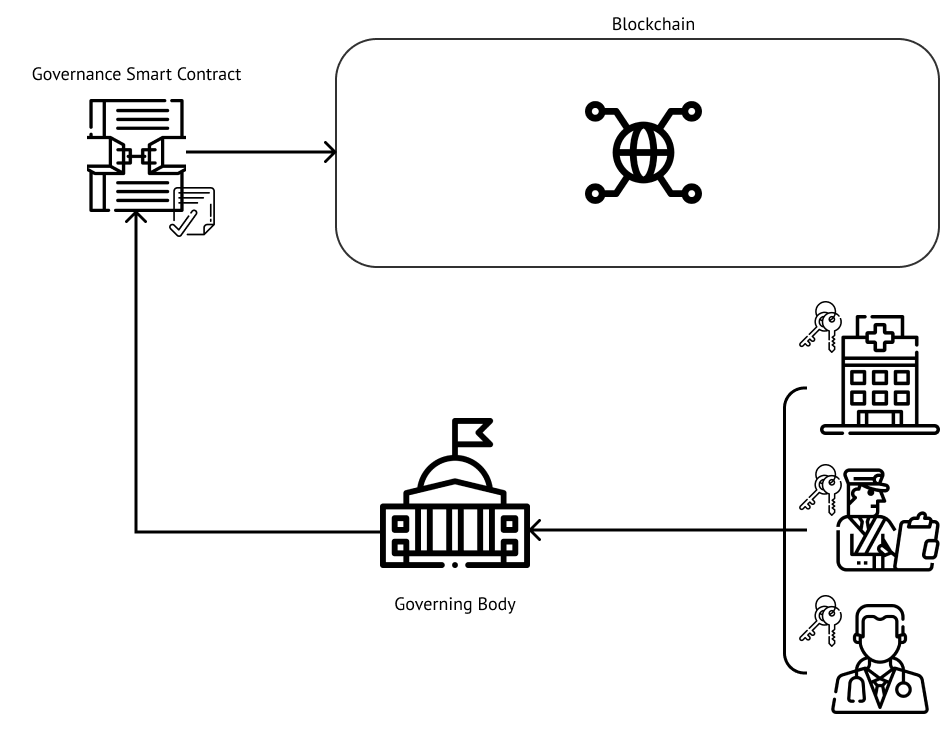}
    \caption{Registration process}
    \label{fig:registration}
\end{figure}

The \ci registration is a process in which \cis are registered and instantiated into the system as depicted in figure \ref{fig:registration}. It assumed that the \gb has deployed the governance smart contract as described in \ref{subsubsec:governance smart contract}.

 The governance contract, retains a mapping data structure that, for each \ci public key, maps the corresponding attributes and can only be altered by the \gb who is responsible for the governance of the system and owner of the contract. The mapping is considered as a trustworthy source of information regarding which issuers are approved by the \gb. If a \ci misbehaves or gets compromised, the \gb can remove the list or revoke access to the \ci in question, making it unable to submit any new transactions.

Before the registration, it is assumed that the \gb has deployed and owns the governance smart contract. Let $(\pk_{i}, \sk_{i})$ be the key pair of each \ci. Also let, $(L_i \cdots L_n)$ the set of attributes which correspond to each \ci.

\begin{algorithm}{
    \caption{\ci registration}\label{prot:ci registration} 
    \procedureblock{}{
    \textbf{\gb} \< \< \textbf{\ci} \\
    \text{Set } P \< \<  (\pk_{i}, \sk_{i}), (L_i \cdots L_n) \\
    \< \sendmessageleft*{\pk_i, (L_i \cdots L_n)} \<  \\
    \< \sendmessageright*{\text{Schnorr protocol}} \<  \\
    \< \sendmessageleft*{} \<  \\
    \text{If } \pk_i \text{ is verified} \\
    S = \sign_{sk_{gb}}(\pk_i, L_i \cdots L_n) \< \< \\
    P \cup (\pk_i, L_i \cdots L_n, S) \< \< \\
    \< \< 
    }
}
\end{algorithm}

The \ci sends their public key $\pk$ along with the set of corresponding attributes $(L_i \cdots \L_n)$. The \gb and the \ci run Schnorr's identification scheme\cite{schnorr89}. Upon the successful run of the protocol, the \gb is convinced that the \ci holds the secret key that was used to generate the public key. The \gb then updates the governance smart contract with the address and attributes received.

\subsection{\vf Registration}
\label{subsec:verifier registration}

\vfs shall be registered into \hygiea by following the same approach as described in section \ref{subsec:certificate issuer registration}. The practical difference between a \vf and a \ci is that \vfs do not issue certificates thus it is not necessary to specify the certificate types data field.

\subsection{Certificate Issuance}
\label{subsection:certificate issuance}
\begin{figure}
    \centering
    \includegraphics[scale=0.3]{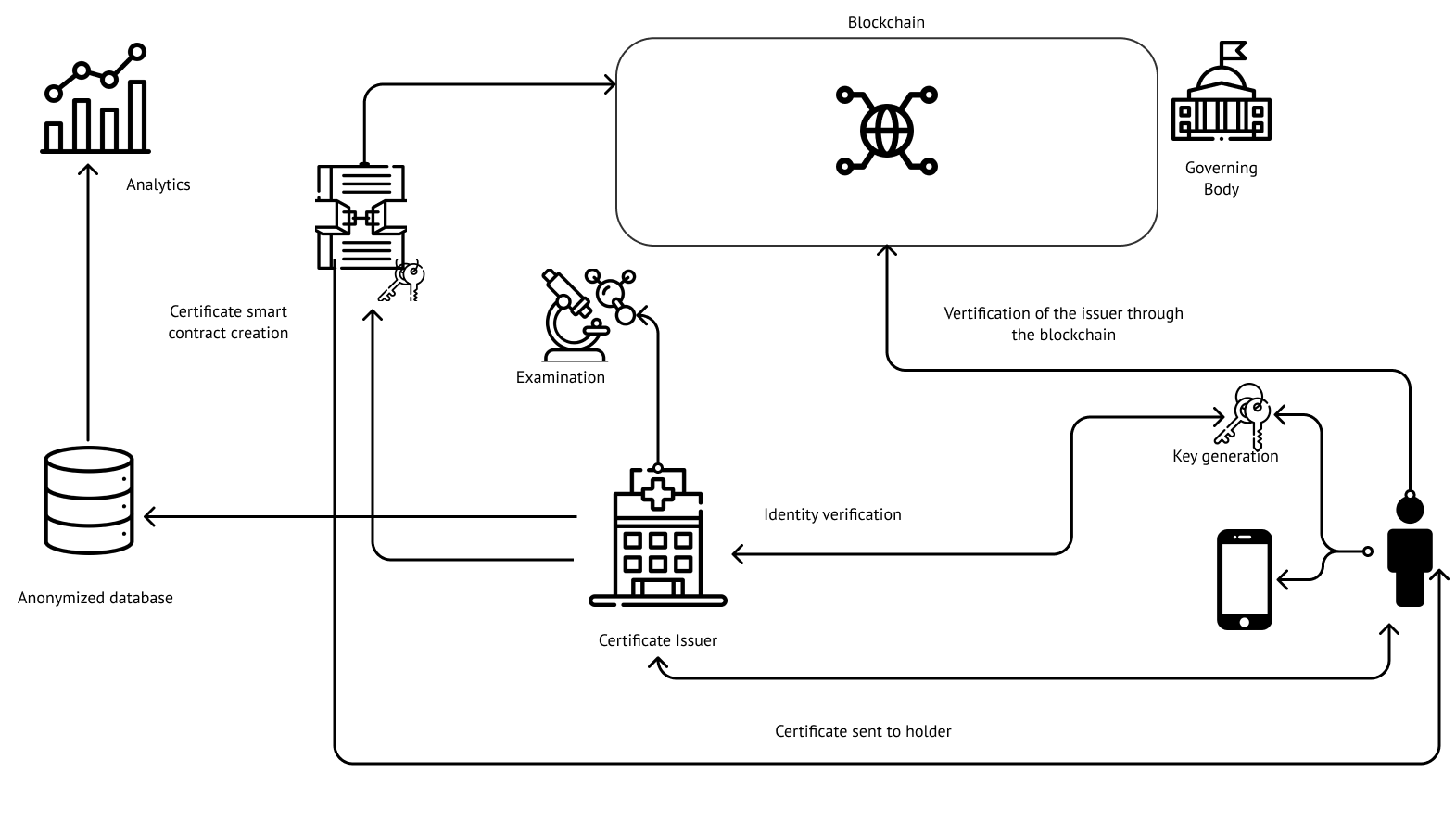}
    \caption{Certificate issuance}
    \label{fig:certificate_issuance}
\end{figure}

Certificate issuance involves the most interactions compared to the rest of the processes in \hygiea. The stakeholders participating are the \ch, the \ci and the \gb. The flow of the process is depicted in Figure \ref{fig:certificate_issuance}. It is assumed that the \gb has deployed the smart contracts on the blockchain and has registered at least one \ci. In addition, the \ci has downloaded the mobile application on his mobile device and has generated his cryptographic keys. The process described here is an indicative example in order to demonstrate the functionalities of \hygiea.

The \ch visits a \ci in order to undertake an examination or get vaccinated. The \ci is required to authenticate the \ch in a hybrid manner. He first verifies the identity of the holder by requesting an identification document(national id, passport, driving license etc.) in order to minimise the possibility of impersonation and bind the physical identity with the certificate. The second phase of authentication is to run Schnorr's identification scheme, similarly to section \ref{subsec:certificate issuer registration}. The purpose of this step is to verify the public key of the \ch. At this point, the \ci should be convinced that the \ci is who he claims to be. 

The holder then proceeds for the medical examination and asks to provide his consent on sharing anonymous metadata with official authorities for further analysis.

As soon as the \ci has all the data required including the test result/vaccination, he then proceeds with the issuance of the corresponding certificate. The certificate, as discussed in \ref{subsubsec:certificate}, is essentially a smart contract that holds the relevant data. The \ci, using the certificate factory, submits a transaction to the blockchain and deploys the certificate contract. 

Once the transaction containing the certificate is submitted to the blockchain, the \gb validates the transaction and updates the state. The address of the certificate is sent to the \ch in order to store it on his mobile wallet. The certificate can then be used for verification.

Algorithm \ref{prot:certificate issuance} shows the process of certificate issuance. Let $(A_i, \cdots, A_n)$ be all the attributes that correspond to the certificate, $(\sk_I, \pk_I)$ be the keypair of the \ci, and $(\sk_H, \pk_H)$ the keypair of the \ch. Let $B_H$ be the identity binding data of the holder.

\begin{algorithm}{
    \caption{Certificate Issuance}\label{prot:certificate issuance} 
    \procedureblock{}{
    \textbf{\ci} \< \< \textbf{\ch} \\
    (\sk_{I}, \pk_{I}) \sample \kgen \secparam \< \< (\sk_{H}, \pk_{H}) \sample \kgen \secparam \\
    \< \< B_H \leftarrow \text{Identity binding data} \\
    \< \sendmessageleft*{\pk_H} \<  \\
	\< \sendmessageright*{\text{Schnorr protocol}} \<  \\
    \< \sendmessageleft*{} \<  \\
    \< \sendmessageleft*{B_H} \<  \\
    O = \sign_{\sk_I}(\pk_H, A_i \cdots A_n, B_H) \\
    C = \text{address}(\pk_H, O) \\
    \< \sendmessageright*{C} \<  \\
    \< \< \text{Store } C \\
    }
}
\end{algorithm}

\subsubsection{Identity binding}
At the certificate issuance step it is essential to have some sort of mechanism to prevent impersonation attacks. We must ensure malicious users are not issued certificates and malicious users should not be allowed to claim a stolen certificate and be verified. Identity binding is the mechanism which binds someone's civil identity (national id, passport, driving license) with the digital identity (public key). One of the main objectives of \hygiea is to preserve the privacy of users and on the same time be resistant to impersonation attacks. In this section we consider a non-exhaustive list of identity binding mechanisms that achieve the various tradeoffs between privacy and security.

The first mechanism we consider is partial personal information binding. This binding mechanism requires that only partial information is included in the certificate smart contract. The exact amount of information depends on the accepted tolerance on collisions that could occur. For example, partial information could be considered as the initials of someone's name, last 4 digits of the national identification document and the full date of birth.

The second mechanism is more aggressive where full personal information is included in the certificate smart contract.  This option minimises the risk of impersonation but sacrifices privacy. Essentially, if a certificate smart contract holds all personal information then the risk of impersonation is the same as the risk of forging a national identification document. This mechanism has been selected by the European Union's DGC.

A third mechanism requires the use of a hash function such as SHA-3\cite{sha3}. All personal information of the user(i.e. name, surname, date of birth etc.) can be the input to the hash function and the output will be a fixed length digest. The digest can be embedded to the certificate smart contract and signed along with the rest of information. For a malicious actor with access to any smart contract there will be no practical way to recover any personal information from the digest or forge it to impersonate the legitimate user.

\subsection{Certificate Verification}
\label{subsubsection:certificate verification}
\begin{figure}
    \centering
    \includegraphics[scale=0.3]{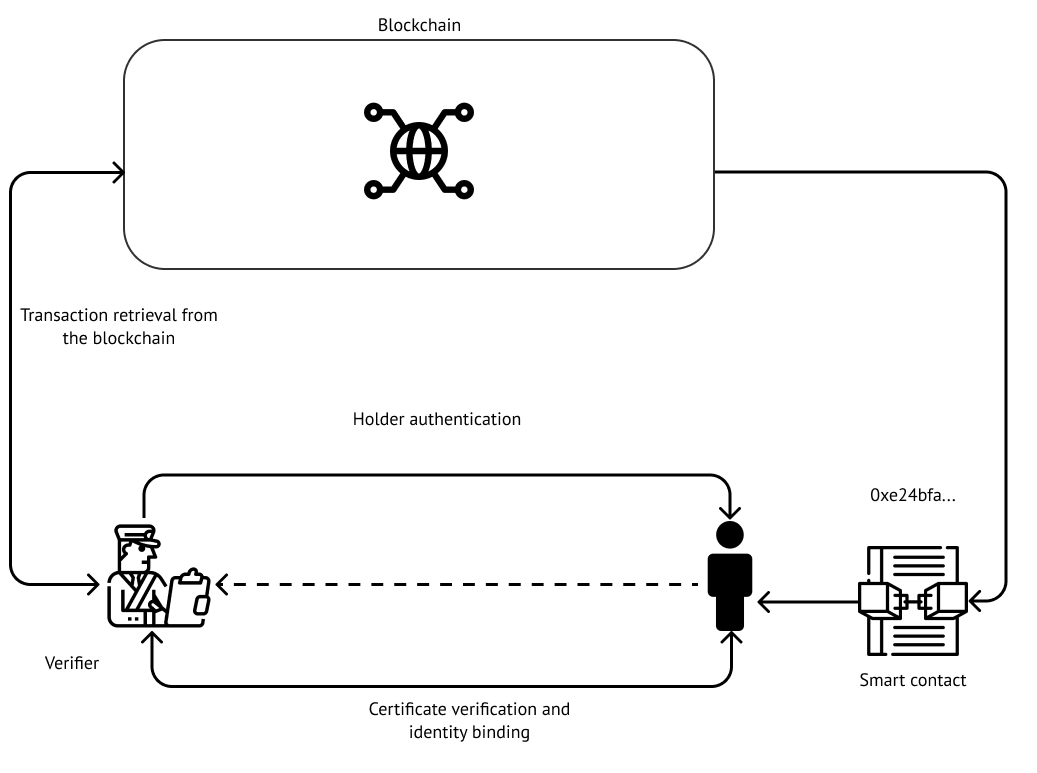}
    \caption{Certificate Verification}
    \label{fig:verification}
\end{figure}

Certificate verification is the process whereby a holder presents certificates to a verifier and the verifier must authenticate the holder and the certificate. Figure \ref{fig:verification} depicts the process. 

To better understand the process consider the following scenario. A citizen who has been issued a certificate visits a hospital or an airport and is required to present a valid vaccination certificate. The citizen using the dedicated mobile application presents the certificate to the verifier. The verifier scans the address of the certificate on the blockchain and checks whether the claimed certificate is valid or not. A verification process involves the following steps:
\begin{itemize}
    \item Check if the public key on the smart contract $\pk_H^\prime$ matches the public key $\pk_H$ sent by the holder.
    \item Check all the relevant attributes of the certificate (see section \ref{subsubsec:certificate}).
    \item Obtain the public key $\pk_I$ of the \ci who signed the certificate.
    \item Verify the signature using the public key of the issuer.
    \item Verify that the certificate is bound to the holder by checking the identifier field.
\end{itemize}

If the above are successful, then the verifier has validated the certificate and ensured that it belongs to the identity that it has been issued to. Algorithm \ref{prot:certificate verification} shows the interactions between the verifier and the certificate holder. Let $C_H$ be the address of the certificate on the blockchain and $B_H$ the binding information of the holder. Let $A_i \cdots A_n$ be the certificate attributes retrieved from the smart contract.

\begin{algorithm}{
    \caption{Certificate Verification}\label{prot:certificate verification} 
    \procedureblock{}{
    \textbf{\vf}  \< \< \textbf{\ch} \\
    (\sk_{V}, \pk_{V}) \sample \kgen \secparam \< \<  (\sk_{H}, \pk_{H}) \sample \kgen \secparam\\
    \< \< B_H \leftarrow \text{Identity binding data} \\
    \< \< C_H \leftarrow \text{Certificate} \\
    \< \sendmessageleft*{B_H, C_H, \pk_H} \<  \\
    \text{Check if } \pk_H \in C_H \\
	\< \sendmessageright*{\text{Schnorr protocol}} \<  \\
    \< \sendmessageleft*{} \<  \\
    \verify_{\pk_I}(\pk_H, A_i \cdots A_n, B_H) \stackrel{?}{=} 1
    }
}
\end{algorithm}

\hygiea supports two modes of verification. The \textit{online} verification mode requires that the verifier has a real-time view of the blockchain which implies that there is network connectivity. In this mode, the verifier apart from verifying the signature of the certificate against the data, he can also check whether the \gb has revoked the certificate. The \textit{offline} verification mode does not real-time access to the blockchain to perform a verification, however, if a certificate has been revoked there will be no way to know. To overcome this limitation in places where network connectivity is not an option, the verifier can keep a relatively recent list of revoked certificates that is updated when possible.

\section{Analytics}
\label{sec:analytics}
\hygiea's second objective is to assist healthcare professionals into better understanding virus spread as well as decision makers towards improved decisions based on accurate data. This can be achieved through analytics derived from data collection and processing during the medical examination or vaccination process. 

For healthcare professionals, such analytics can assist into better understanding the behaviour of the virus, the potential mutations and ultimately improve the efficiency of vaccines and medicine. The ability to predict potential areas of infection can be used for better preparation of healthcare service providers. Analytics can also be derived after vaccination campaigns in order to assess the effectiveness of the campaign and potentially improve the process.

\begin{figure}
    \centering
    \includegraphics[scale=0.8]{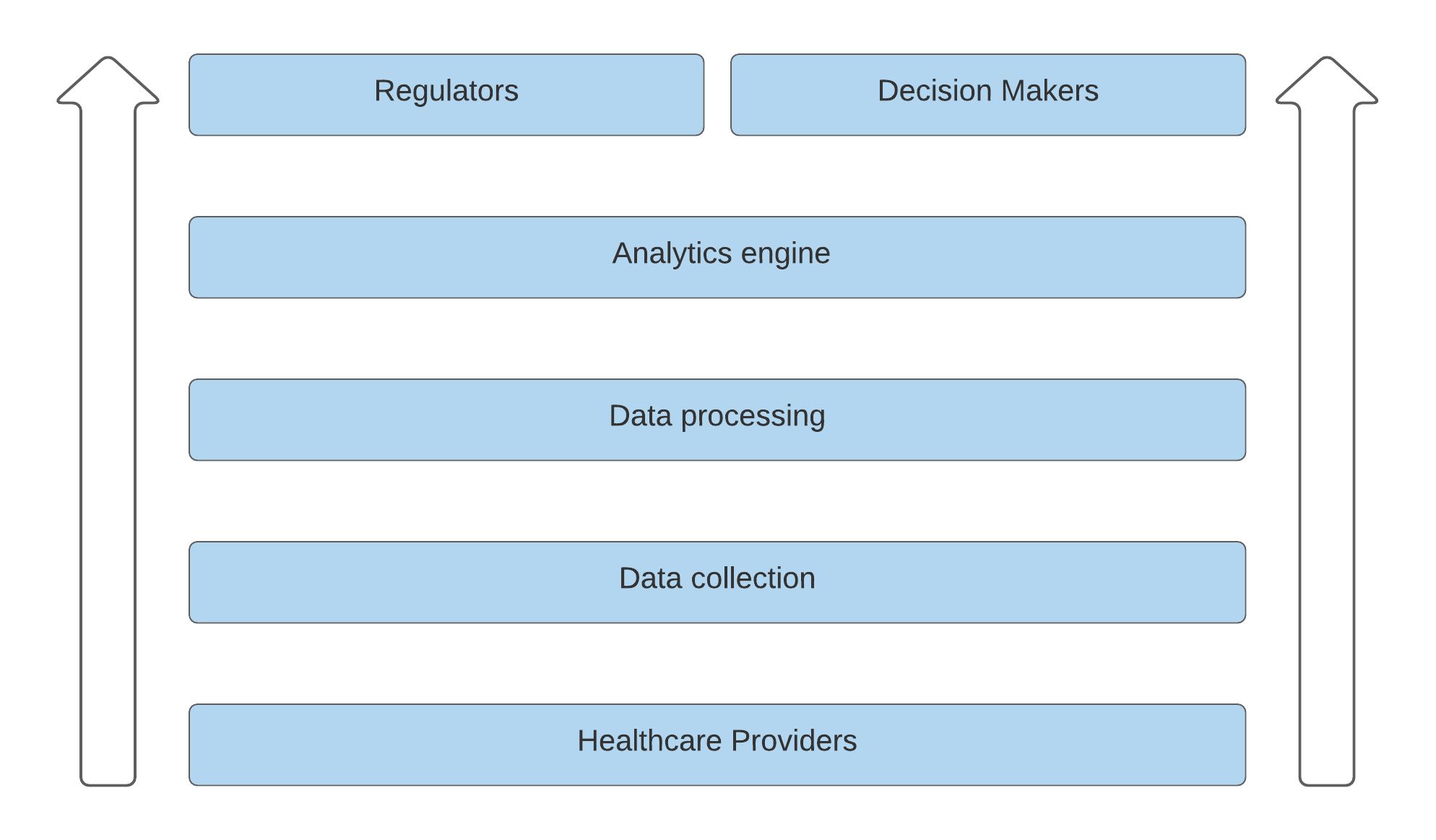}
    \caption{Analytics flow}
    \label{fig:analytics flow}
\end{figure}

For decision makers in governments or regulatory bodies, real-time analytics can greatly influence the quality of the decisions taken to contain the virus. It may also serve to enforce more effective lockdowns and measure the impact with better accuracy. In addition, analytics can improve the allocation of both human resources and medical equipment in order to prevent over-crowded hospitals. Figure \ref{fig:analytics flow} depicts the flow of information from the healthcare providers to the analytics engine.

\subsection{Data collection}
\label{subsec:data collection}
The design of \hygiea covers the data collection aspect during the certificate issuance process where the citizen may opt in for anonymous data collection. Certificate issuers who perform examinations or vaccinations will be given access to a web application that will collect data in a structured way in order to produce real-time analytics. The process is straight-forward. During an examination or vaccination, the certificate issuer will ask for the consent of the citizen to provide anonymised data. Upon confirmation of the citizen's identity, the issuer will fill in a certain form and will submit the data into a dedicated database.

Below we list the Covid-19 related data fields that have been identified by the Cyprus Institute of Neurology \& Genetics\cite{CING} who served as a subject matter expert in this work:
\begin{itemize}
	\item Fever or chills
	\item Cough
	\item Shortness of breath or difficulty breathing
	\item Fatigue
	\item Muscle or body aches
	\item Headache
	\item Loss of smell
	\item Loss of taste
	\item Sore throat
	\item Congestion or runny nose
	\item Nausea or vomiting
	\item Diarrhea
	\item Age
	\item Geolocation
	\item Travel history
	\item Past infections
\end{itemize}

\subsection{Analytics engine}
\begin{figure}
    \centering
    \includegraphics[scale=0.5]{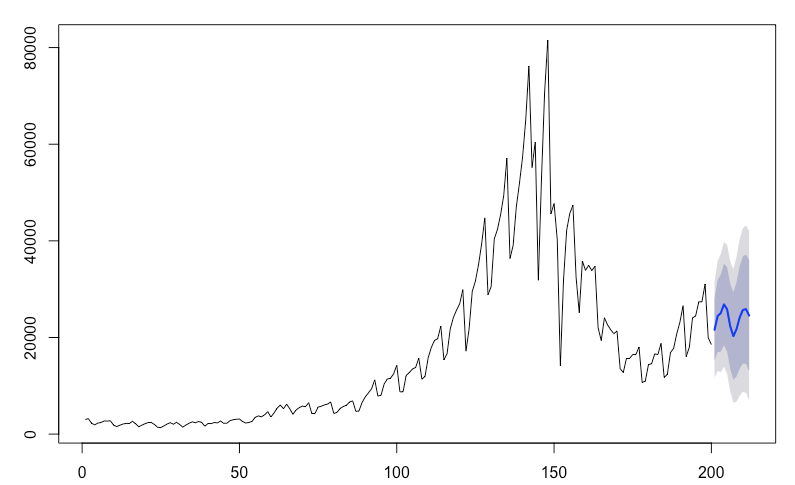}
    \caption{Covid-19 cases forecast}
    \label{fig:forecasted_covid_cases}
\end{figure}

\begin{figure}
    \centering
    \includegraphics[scale=0.5]{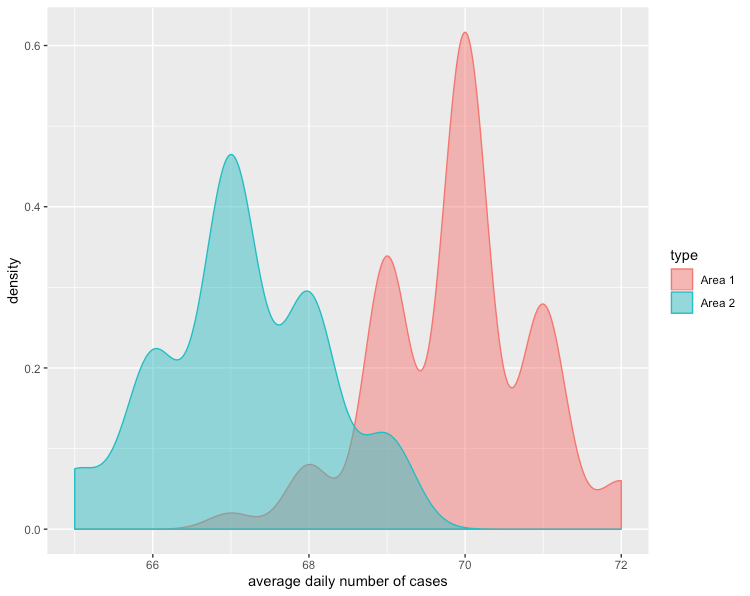}
    \caption{Example virtualisation of how the system can automatically detect and plot differences in the total number of cases across different areas}
    \label{fig:example virtualization}
\end{figure}

\hygiea supports general analytics related to the pandemic spread, vaccinations and tests. Some examples of the supported analytics are the total number of cases, distribution of cases across regions or the performance of vaccination and testing initiatives. 
In a such a system where there is a significant flow of information it is required that an automatic inference mechanism is in place. This mechanism constantly analyses the data and detects suspicious patterns in space and time that might require closer attention. Examples include the deviations in the rate of exposure between different areas or across time. The extraction of such patterns is done through the use of statistical methodologies and automated machine learning. Figure \ref{fig:example virtualization} depicts an example virtualisation of how the system detects and plots differences in the total number of cases across different areas. Figure \ref{fig:forecasted_covid_cases} depicts an example of a Covid-19 case forecast.

In addition to inferencing, \hygiea can be used by experts for automatic estimation and forecasting. It is possible to derive parameters, such as the infection rate using the susceptible-infected-removed (SIR) or other modifications of it such as \cite{hethcote2000mathematics}. More sophisticated forecasting algorithms based on Bayesian structural time series \cite{Bayesiantimeseries} models can be used in order to forecast transmission trends.

Finally, \hygiea automatically plots the distribution of symptoms, and can give early insights which help the authorities understand which symptoms are most prominent, in order to inform policy.

\subsection{Data quality checks}
One of the key challenges of fighting pandemics has been the accurate collection and recording of data. \hygiea implements automated data checks, which ensure that there are no significant errors in the data. As more data is collected, then the platform detects any outliers or unusual data points which need special attention. When the platform runs analyses, it conducts these analyses both with and without the potentially problematic data points, and it automatically compares and contrasts the results. This allows scenario analysis of multiple confidence levels. This is based on a blend of anomaly and outlier detection algorithms, which combines isolation forests\cite{isolationforest}, one-class SVM\cite{svmanomalydetection}, and local outlier factor\cite{localoutlierprobabilities}.

\subsection{Compliance with government policies}
The key stakeholders (e.g. governments or hospitals) are able to add smart contracts that track the compliance with key government policies. For example, in the UK, minorities were hit harder by COVID-19. The UK Office for National Statistics recently released a comment \cite{nationalstatistics} regarding that. \hygiea automatically detects such instances and issues warnings on cases like differences in the number of people inflicted by the virus based on their demographics. Furthermore, it detects differences between rates of vaccination or testing and gender or ethnicity. Finally, it includes methodologies like the sequential probability ratio test to detect significant increases in the number of cases.

\section{Conclusions}
In this work we have designed and implemented a blockchain-based platform, namely \hygiea, that addresses the Covid-19 certificate problem. Governments deploying \hygiea are ensured it is sufficiently hard for citizens to forge Covid-19 certificates and on the same time citizens are ensured on their privacy is preserved. 

 The system covers all the steps involved from the governance, issuance and verification of Covid-19 related certificates. The platform has well-specified roles for all stakeholders and utilises smart contracts to define its business logic. The governance of the system is implemented via smart contracts and is coordinated by a trusted entity in the system. Smart contracts and the blockchain offer transparency, security and auditability. User data and interactions are designed with privacy in mind and users are in full control over to whom they disclose their data.

\hygiea's secondary objective is to provide tools to experts and decision makers to better understand the spread of the disease. In this work we have shown how our system can be used to collect anonymous data and process them in order to generate valuable inferences and forecasts that will help experts to better understand the disease and decision makers towards improved decisions related to the pandemic.

\clearpage
\bibliographystyle{unsrt}
\bibliography{bibtex}

\begin{thebibliography}{10}

\bibitem{JohnHopkinsCovid}
{COVID-19 Dashboard by the Center for Systems Science and Engineering (CSSE) at
  Johns Hopkins University (JHU)}.
\newblock \url{https://coronavirus.jhu.edu/map.html}.
\newblock Accessed: 2021-03-20.

\bibitem{Halpin2020}
Harry Halpin.
\newblock Vision: A critique of immunity passports and w3c decentralized
  identifiers.
\newblock In Thyla van~der Merwe, Chris Mitchell, and Maryam Mehrnezhad,
  editors, {\em Security Standardisation Research}, pages 148--168, Cham, 2020.
  Springer International Publishing.

\bibitem{Voo2020}
Teck~Chuan Voo, Hannah Clapham, and Clarence~C Tam.
\newblock {Ethical Implementation of Immunity Passports During the COVID-19
  Pandemic}.
\newblock {\em The Journal of Infectious Diseases}, 222(5):715--718, 06 2020.

\bibitem{VaccinePassportMonitor}
{International monitor: vaccine passports and COVID status apps}.
\newblock
  \url{https://www.adalovelaceinstitute.org/project/international-monitor-vaccine-passports-covid-status-apps/}.
\newblock Accessed: 2021-04-08.

\bibitem{EUDigitalGreenCertificate}
{Coronavirus: Commission proposes a Digital Green Certificate}.
\newblock
  \url{https://ec.europa.eu/commission/presscorner/detail/en/ip_21_1181}.
\newblock Accessed: 2021-03-17.

\bibitem{Hasan2020}
H.~R. {Hasan}, K.~{Salah}, R.~{Jayaraman}, J.~{Arshad}, I.~{Yaqoob}, M.~{Omar},
  and S.~{Ellahham}.
\newblock Blockchain-based solution for covid-19 digital medical passports and
  immunity certificates.
\newblock {\em IEEE Access}, 8:222093--222108, 2020.

\bibitem{Eisenstadt2020}
Marc Eisenstadt, Manoharan Ramachandran, Niaz Chowdhury, Allan Third, and John
  Domingue.
\newblock Covid-19 antibody test/vaccination certification: There’s an app
  for that.
\newblock {\em IEEE Open Journal of Engineering in Medicine and Biology},
  1:148–155, 2020.

\bibitem{VerifiableCredentials}
{Verifiable Credentials W3C}.
\newblock \url{https://www.w3.org/TR/vc-data-model/}.
\newblock Accessed: 2021-03-31.

\bibitem{Solid}
{Solid Project}.
\newblock \url{https://solidproject.org/}.
\newblock Accessed: 2021-03-31.

\bibitem{Untung2020}
Untung Rahardja, Ankur~Singh Bist, Marviola Hardini, Qurotul Aini, and
  Eka~Purnama Harahap.
\newblock Authentication of covid-19 patient certification with blockchain
  protocol.
\newblock {\em International Journal of Advanced Science and Technology},
  29(8s):4015 -- 4024, Apr. 2020.

\bibitem{secureabc2020}
Chris Hicks, David Butler, Carsten Maple, and Jon Crowcroft.
\newblock Secureabc: Secure antibody certificates for covid-19, 2020.

\bibitem{aydar2020blockchain}
Mehmet Aydar, Serkan Ayvaz, and Salih~Cemil Cetin.
\newblock Towards a blockchain based digital identity verification, record
  attestation and record sharing system, 2020.

\bibitem{Ahmad2020}
Raja~Wasim Ahmad, Khaled Salah, Raja Jayaraman, Ibrar Yaqoob, Samer Ellahham,
  and Mohammed Omar.
\newblock {Blockchain and COVID-19 Pandemic: Applications and Challenges}.
\newblock 9 2020.

\bibitem{hernandez2021sharing}
Jos{\'e}~Luis Hern{\'a}ndez-Ramos, Georgios Karopoulos, Dimitris Geneiatakis,
  Tania Martin, Georgios Kambourakis, and Igor~Nai Fovino.
\newblock Sharing pandemic vaccination certificates through blockchain: Case
  study and performance evaluation.
\newblock {\em arXiv preprint arXiv:2101.04575}, 2021.

\bibitem{Hyperledger}
{Hyperledger}.
\newblock \url{https://www.hyperledger.org//}.
\newblock Accessed: 2021-03-31.

\bibitem{chaudhari2021framework}
Sarang Chaudhari, Michael Clear, and Hitesh Tewari.
\newblock Framework for a dlt based covid-19 passport, 2021.

\bibitem{Europol}
{EUROPOL WARNING ON THE ILLICIT SALE OF FALSE NEGATIVE COVID-19 TEST
  CERTIFICATES}.
\newblock
  \url{https://www.europol.europa.eu/newsroom/news/europol-warning-illicit-sale-of-false-negative-covid-19-test-certificates}.
\newblock Accessed: 2021-04-12.

\bibitem{bitcoin2009}
Satoshi Nakamoto.
\newblock Bitcoin: A peer-to-peer electronic cash system.
\newblock {\em Cryptography Mailing list at https://metzdowd.com}, 03 2009.

\bibitem{wood2014ethereum}
Gavin Wood et~al.
\newblock Ethereum: A secure decentralised generalised transaction ledger.
\newblock {\em Ethereum project yellow paper}, 151(2014):1--32, 2014.

\bibitem{Solidity}
{Solidity}.
\newblock \url{ https://docs.soliditylang.org/}.
\newblock Accessed: 2021-03-20.

\bibitem{BlockchainInHealthcare}
Anton Hasselgren, Katina Kralevska, Danilo Gligoroski, Sindre~A. Pedersen, and
  Arild Faxvaag.
\newblock Blockchain in healthcare and health sciences—a scoping review.
\newblock {\em International Journal of Medical Informatics}, 134:104040, 2020.

\bibitem{secp256k1}
{ECDSA secp256k1 parameters}.
\newblock \url{http://www.secg.org/sec2-v2.pdf}.
\newblock Accessed: 2021-05-18.

\bibitem{DSA}
{Digital Signature Standard}.
\newblock \url{https://nvlpubs.nist.gov/nistpubs/FIPS/NIST.FIPS.186-4.pdf}.
\newblock Accessed: 2021-05-21.

\bibitem{schnorr89}
C.~P. Schnorr.
\newblock Efficient identification and signatures for smart cards.
\newblock In Gilles Brassard, editor, {\em Advances in Cryptology --- CRYPTO'
  89 Proceedings}, pages 239--252, New York, NY, 1990. Springer New York.

\bibitem{designpatterns1995}
Erich Gamma, Richard Helm, Ralph Johnson, and John Vlissides.
\newblock {\em Design Patterns: Elements of Reusable Object-Oriented Software}.
\newblock Addison-Wesley Longman Publishing Co., Inc., USA, 1995.

\bibitem{sha3}
{Keccak implementation overview}.
\newblock \url{https://keccak.team/files/Keccak-implementation-3.2.pdf}.
\newblock Accessed: 2021-05-20.

\bibitem{CING}
{Cyprus Institute of Neurology and Genetics}.
\newblock \url{https://www.cing.ac.cy/}.
\newblock Accessed: 2021-05-28.

\bibitem{hethcote2000mathematics}
Herbert~W Hethcote.
\newblock The mathematics of infectious diseases.
\newblock {\em SIAM review}, 42(4):599--653, 2000.

\bibitem{Bayesiantimeseries}
Kay~H. Brodersen, Fabian Gallusser, Jim Koehler, Nicolas Remy, and Steven~L.
  Scott.
\newblock Inferring causal impact using bayesian structural time-series models.
\newblock {\em Annals of Applied Statistics}, 9:247--274, 2015.

\bibitem{isolationforest}
Fei~Tony Liu, Kai~Ming Ting, and Zhi-Hua Zhou.
\newblock Isolation forest.
\newblock In {\em 2008 Eighth IEEE International Conference on Data Mining},
  pages 413--422, 2008.

\bibitem{svmanomalydetection}
K.~Li, Houkuan Huang, S.~Tian, and W.~Xu.
\newblock Improving one-class svm for anomaly detection.
\newblock {\em Proceedings of the 2003 International Conference on Machine
  Learning and Cybernetics (IEEE Cat. No.03EX693)}, 5:3077--3081 Vol.5, 2003.

\bibitem{localoutlierprobabilities}
{Hans Peter} Kriegel, Peer Kr{\"o}ger, Erich Schubert, and Arthur Zimek.
\newblock Loop: Local outlier probabilities.
\newblock In {\em Proceedings of the 18th ACM conference on Information and
  knowledge management}, pages 1649--1652, United States, December 2009.
  Association for Computing Machinery.
\newblock ACM 18th International Conference on Information and Knowledge
  Management ; Conference date: 02-11-2009 Through 06-11-2009.

\bibitem{nationalstatistics}
{UK Office for National Statistics}.
\newblock
  \url{https://www.ons.gov.uk/peoplepopulationandcommunity/healthandsocialcare/conditionsanddiseases/articles/whyhaveblackandsouthasianpeoplebeenhithardestbycovid19/2020-12-14}.
\newblock Accessed: 2021-05-31.

\end{thebibliography}

\end{document}